# *From Algorithmic Marginalization to AI-Mediated Re-Centering: Can Culturally Grounded AI Bring Hakka Language and Culture Back into Mainstream Society?*

Chen-Chi Chang

Department of Cultural Creativity and Digital Marketing, College of Hakka Studies, National United University, Taiwan

**Publication Note:** This Research Note is published in the *GHAS Newsletter* of the Consortium of Global Hakka Studies 2026 (GHAS, 2026).

## Abstract

As large language models increasingly mediate writing, translation, information retrieval, and education, the technological support available to a language may influence its position in contemporary social life. For minority and minoritized languages, this raises a dual problem: inadequate functionality can encourage movement toward dominant languages, while apparently fluent assistance can normalize culturally distinctive expression. This conceptual Research Note develops a framework connecting algorithmic cultural marginalization, cultural grounding, and AI-mediated cultural re-centering. Algorithmic cultural marginalization describes the interaction of representational asymmetry, functional exclusion, homogenizing transformation, and recursive feedback. Cultural grounding identifies interventions across resources, models, services, and governance. Cultural re-centering concerns changes in accessibility, actual language use, public visibility, participation, and the circulation of new knowledge. Drawing on research on language shift and language technologies, the paper uses Hakka in Taiwan as an illustrative case and develops eight propositions for subsequent empirical investigation. Its central argument is that expanded AI access should be assessed alongside the retention of linguistic variation and community authority. The framework distinguishes technological availability from social use, and factual accuracy from cultural fidelity, without assuming that AI can substitute for intergenerational transmission. It provides a basis for investigating whether minority-language AI expands meaningful domains of use or reproduces linguistic marginalization through convenience and standardization.

## 1 Introduction

Large language models (LLMs) increasingly mediate activities through which language is used to obtain knowledge, complete tasks, and participate in public life. Their significance for minority languages consequently extends beyond the accuracy of individual translations or generated sentences. When writing assistance, information retrieval, educational support, and conversational services are substantially more effective in one language than another, technological infrastructures can alter the practical conditions of language choice. A language may retain cultural recognition while becoming less usable in the activities that organize contemporary digital life.

Research on language maintenance has long located language shift within changing social arrangements rather than treating it simply as an aggregate of individual preferences. Fishman (1991) emphasized intergenerational transmission and the distribution of language use across families, communities, education, and institutions. Eisenlohr (2004) subsequently demonstrated that communication technologies can reorganize cultural authority and the boundaries of speech communities. Kornai (2013) extended this discussion to digital environments, distinguishing the availability of archived linguistic material from the capacity of a language to sustain digitally mediated communication. These perspectives establish that technological access matters because it can change where, by whom, and for what purposes a language is used.

Generative AI introduces a further complication. Unlike an archive or a communication channel, an LLM actively selects, reformulates, translates, and produces expression. It therefore affects both access to linguistic resources and the forms through which those resources are encountered. Unequal support can discourage minority-language use, but improved support can also privilege a narrow range of varieties, genres, or cultural perspectives. Technical inclusion and the retention of linguistic diversity are related concerns, yet neither can be inferred from the other.

Existing scholarship provides evidence for several components of this problem. Joshi et al. (2020) document the unequal distribution of natural language processing resources across languages. Blodgett et al. (2020) locate language technology within social hierarchies and emphasize the need to specify whose interests are served by particular definitions of performance and bias. Ryan et al. (2024) show that alignment can produce disparities across English dialects and global viewpoints. Research on AI-assisted production identifies additional risks: individual creative outputs can improve while becoming collectively more similar (Doshi & Hauser, 2024), and writing suggestions can shift expression toward Western cultural patterns (Agarwal, Naaman, & Vashistha, 2025).

Sourati et al. (2026) further demonstrate that semantic preservation does not guarantee the preservation of linguistic variation. Across seven datasets comprising more than 880,000 texts, LLM-assisted rewriting reduced variation in writing complexity and weakened identity-related linguistic cues while retaining core content. These findings do not establish an equivalent effect for Hakka or other minority languages. They nevertheless motivate a theoretically consequential question: if AI systems repeatedly normalize the expression through which cultural difference is communicated, what happens to the social position and internal diversity of less represented languages?

This paper addresses that question through a conceptual framework linking algorithmic cultural marginalization, cultural grounding, and AI-mediated cultural re-centering. It asks under what sociotechnical conditions AI may reproduce linguistic marginalization, and under what conditions it may expand meaningful participation through a minority language. The argument is developed through an interpretive synthesis of selected scholarship rather than a systematic literature review or an empirical evaluation of a particular system. Hakka in Taiwan serves as an illustrative case because research on its language shift demonstrates how changes in social environments can affect language retention (Jan et al., 2016; Wang & Peng, 2018).

The paper contributes an account of how model behavior, user adaptation, and the subsequent production of linguistic resources may interact. It also specifies cultural grounding as a design and governance intervention, distinct from the observed social outcomes used to assess re-centering. Eight propositions articulate relationships that require empirical testing. The guiding principle is mainstreaming without homogenization: expanding the contexts in which a minority language can be used while preserving legitimate alternatives within that language and the authority of communities to determine their use.

## 2 Language Shift and Algorithmic Cultural Marginalization

### 2.1 From digital access to algorithmic mediation

The relationship between technological support and language vitality is best understood through domains of use. A language need not lose all of its speakers to lose access to important social activities. When another language becomes the default for education, institutional communication, professional work, or media consumption, the relative usefulness of the minority language can decline even where speakers continue to value it. Fishman’s (1991) account of language maintenance thus directs attention to the environments in which language practices are reproduced.

Digital technologies create comparable pressures, although their consequences depend on how they are incorporated into social life. Eisenlohr (2004) cautions against treating electronic mediation as the neutral preservation of an already established community. Kornai (2013) likewise distinguishes digital heritage from digital vitality: documentation may survive without enabling sustained communication among speakers. Generative AI extends these concerns because it mediates the completion of tasks and the production of new linguistic material, rather than merely providing a location in which existing materials can be stored.

This distinction helps explain why a list of supported languages is an inadequate measure of inclusion. A model may nominally generate a minority language while producing unstable orthography, inaccurate transcription, dominant-language substitutions, or culturally inappropriate responses. If users must switch languages to complete a task reliably, nominal availability does not provide functional parity. However, language switching should not automatically be interpreted as marginalization. Multilingual practices can be intentional, culturally meaningful, and useful. The relevant problem is the additional switching attributable to unequal technological support when users would otherwise prefer to continue in the minority language.

## 2.2 Representation and functional exclusion

Algorithmic cultural marginalization is defined here as a cumulative sociotechnical process through which unequal linguistic representation and AI-mediated practices reduce a minority language's visibility, practical usability, distinctiveness, or capacity to generate and circulate new knowledge. The construct combines four mechanisms: representational asymmetry, functional exclusion, homogenizing transformation, and recursive marginalization. These mechanisms can operate simultaneously; they should not be interpreted as a universal sequence that every language follows.

Representational asymmetry concerns unequal coverage across languages and within them. Corpus volume is only one aspect of the problem. A relatively large collection may disproportionately represent standardized writing, institutional documents, a single dialect, or a small number of producers. Oral interaction, local vocabulary, age-related variation, and everyday genres may remain poorly represented. Aggregate improvements in a language-level benchmark can therefore coexist with persistent disadvantages for particular varieties or communities. Joshi et al. (2020) establish the broader inequality of technological support, while Blodgett et al. (2020) explain why the selection of data and evaluation criteria has social consequences.

Functional exclusion concerns the tasks that users can successfully complete through a language. Speech recognition, translation, conversational interaction, and knowledge retrieval impose different requirements, and competence in one does not establish competence in the others. Inferior performance can raise interaction costs through repeated corrections, additional explanations, or movement to a dominant language. The proposed effect is behavioral: uneven support may change language choice even when cultural identification remains stable. This extends the domain perspective on language shift to the practical affordances of AI systems, while leaving the magnitude and duration of the effect open to investigation.

*P1. Greater imbalance in the representation of languages and varieties across training, alignment, retrieval, and evaluation resources will be associated with larger gaps in task performance and linguistic-variety fidelity.*

*P2. Among users able and initially willing to use a minority language, poorer AI task performance in that language will increase switching to a dominant language, after accounting for language proficiency, task requirements, and prior preferences.*

P1 concerns resource and performance differences; P2 concerns user responses to those differences. Keeping these relationships separate avoids defining marginalization merely through the same representational inequalities proposed as its causes. A system can have an uneven resource base without every user abandoning the language, while a small performance gap may become consequential in tasks for which reliability is particularly important.

### 2.3 Homogenization and recursive feedback

Homogenizing transformation concerns the normalization of distinctive linguistic and cultural features during generation, rewriting, translation, or interaction. Relevant features include dialectal vocabulary, discourse conventions, oral forms, kinship terms, culturally situated metaphors, politeness practices, and narrative patterns. These features are not interchangeable with propositional meaning. An output can communicate approximately the same factual content while changing the social relationships, locality, or cultural positioning conveyed by its expression.

The existing evidence supports this distinction but also establishes limits to its generalization. Doshi and Hauser (2024) identify a tension between individual creativity and collective similarity; Agarwal et al. (2025) demonstrate cultural convergence in AI-assisted writing; and Sourati et al. (2026) document reductions in linguistic variation during rewriting. Ryan et al. (2024) show that alignment can redistribute representational advantage, while Wang et al. (2025) find that using LLMs as substitutes for human participants can flatten identity-group representations. These studies examine different tasks and populations. They should be treated as evidence for plausible mechanisms, rather than as direct proof of minority-language decline.

Recursive marginalization concerns the reproduction of these disadvantages through subsequent behavior and resource formation. One possible process operates through use: weak support encourages switching, which reduces minority-language interaction and potentially the production of new human-authored material. Another operates through synthetic content: normalized model outputs may be reused in educational materials, public communication, or later training resources. Guo et al. (2024) demonstrate declining lexical, syntactic, and semantic diversity under recursive training on synthetic text. This provides evidence for a model-level feedback process, but its relationship to community language practices remains an empirical question.

The two processes may reinforce one another, but this is not inevitable. Human correction, changes in institutional support, diversified data collection, and alternative technologies may interrupt them. It is also possible for AI assistance to increase the production of minority-language content without substantially narrowing variation. The framework therefore treats feedback as conditional on the composition of new resources, the extent of model influence, and the opportunities available for independent human expression.

*P3. AI rewriting or generation that privileges dominant linguistic norms will reduce dialectal and cultural-marker variation relative to human-authored source material, even where semantic accuracy remains comparable.*

*P4. Over time, reduced human minority-language production and greater reuse of homogenized synthetic material will be associated with declining coverage of linguistic variation in subsequent resources, unless compensating human contributions or governance interventions interrupt the feedback process.*

## 3 Cultural Grounding and AI Mediated Cultural Re Centering

### 3.1 Grounding across resources models services and governance

If marginalization can emerge through resource selection, model behavior, service design, and user adaptation, increasing corpus size alone offers an incomplete response. Additional data can reproduce an existing imbalance among varieties. Retrieval can supply accurate cultural facts without resolving inappropriate framing. A technically capable model can remain inaccessible to intended users, and participation can expand without giving communities authority over representation or reuse. Cultural grounding addresses these interdependencies rather than treating cultural accuracy as an isolated model attribute.

The term is defined here as the deliberate embedding of linguistic variation, situated cultural knowledge, provenance, practical task support, and community authority across the resource, model, service, and governance layers of an AI system. This definition builds on, rather than replaces, technical approaches to grounding. Lertvittayakumjorn et al. (2025) distinguish propositional cultural knowledge from open-ended cultural familiarity. Their experiments show that search grounding can improve factual benchmark performance without corresponding improvements in judged cultural familiarity, and can increase stereotypical judgments. The implication is that the availability of retrieved cultural information does not itself establish cultural fidelity.

Resource grounding concerns the composition and contextualization of accessible materials. Coverage should reflect relevant variation across dialects, genres, modalities, localities, generations, and cultural domains. Provenance should make it possible to distinguish the location, period, speaker community, and circumstances to which a statement applies. These requirements do not imply a complete or singular representation of a culture. They make plurality and uncertainty available for interpretation, including cases in which sources conflict or practices differ across communities.

Model grounding concerns whether such distinctions influence generated responses. Fine-tuning, retrieval-augmented generation, dialect-aware retrieval, knowledge graphs, and linguistic conditioning are possible mechanisms, but no architecture should be presumed culturally grounded solely because it uses one of them. Evaluation must establish whether the system actually retains requested varieties, uses relevant evidence, distinguishes situated practices from general claims, and communicates uncertainty appropriately. A general improvement in fluency is insufficient if it is achieved by replacing legitimate alternatives with the model’s preferred standard.

Service grounding concerns the design conditions that make minority-language functionality available for meaningful tasks. Speech interfaces, manageable interaction costs, accessible explanations, and connections between translation, retrieval, and production can support use beyond cultural display. The distinction from re-centering is analytical: service grounding concerns implemented capabilities and design provisions, whereas re-centering concerns observed changes in access, use, and participation. A service may be well designed yet attract little sustained use because of institutional, demographic, or economic constraints.

Governance grounding concerns who can authorize collection, define acceptable use, identify consequential errors, correct representations, and determine whether data or outputs may be reused. Bird (2020) warns that language technologies can reproduce colonial relationships when languages are treated as extractable resources. Blodgett et al. (2020) similarly call for attention to affected communities and the power relations of technical development. Community participation must therefore extend beyond supplying data or validating finished products. It should include consequential authority over design and correction, while recognizing that communities contain differing interests and unequal opportunities to participate.

The four layers are complementary interventions rather than interchangeable indicators of a single technical capability. Resource diversity may remain unused by a model; model fidelity may not translate into usable services; and participation may remain consultative without enforceable decision rights. Their contribution should consequently be evaluated both individually and in combination.

*P5. Systems with stronger resource and model grounding will retain more dialectal and culturally situated features than systems with comparable semantic performance but weaker grounding.*

*P6. At comparable levels of resource and model grounding, meaningful community authority over evaluation and correction will improve the identification and remediation of culturally consequential errors.*

### 3.2 Re centering as an observed social outcome

AI-mediated cultural re-centering refers to an expansion in the positions through which a marginalized language can be encountered, used, and reproduced within digital society. It concerns accessibility, functional use, public visibility, participation, and generative circulation. Re-centering does not require a minority language to displace a dominant language numerically. Its object is a change in functional position: speakers and learners gain opportunities to obtain knowledge, complete tasks, communicate, and create through the language.

Accessibility concerns entry into minority-language knowledge and interaction. Heritage learners or people with limited proficiency may approach through a dominant language while encountering minority-language terminology, audio, or situated explanations. Such access can broaden participation, but exposure alone does not demonstrate language vitality. Functional use concerns whether people actually complete meaningful tasks through the minority language and continue to do so beyond initial experimentation. This requires observations of behavior, rather than reliance on the existence of an interface or self-reported interest.

Public visibility concerns the presence of minority-language knowledge within AI-mediated encounters, including whether generated answers acknowledge and accurately contextualize relevant sources. A digital archive can remain available yet be poorly represented in the systems through which users search for knowledge. Participation concerns opportunities for users to contribute, correct, teach, translate, and influence representations. Generative circulation concerns the production and subsequent use of new linguistic material across everyday communication, education, and cultural activity. It is analytically distinct from participation because a system may solicit contributions without enabling their meaningful circulation.

These dimensions constitute an assessment profile rather than a sequence of stages. Increased visibility may not produce sustained use, and additional content may not reflect broader participation. Their interpretation also depends on whether contributions preserve internal variation and whether communities retain authority over how they are used. An increase in machine-generated text is therefore an inadequate proxy for re-centering. The relevant evidence concerns who participates, which tasks become possible, which varieties remain available, and how new material enters subsequent social and technological practices.

*P7. Cultural grounding will be associated with increased minority-language access, successful task completion, public visibility, participation, and circulation of new content, with these relationships mediated or constrained by user capabilities and institutional conditions.*

### 3.3 Mainstreaming without homogenization

The relationship between inclusion and diversity creates a boundary condition for interpreting re-centering. An AI service can expand exposure and increase content volume while systematically privileging one variety or external cultural perspective. Such expansion can be useful for particular purposes, yet it does not establish that the language's internal diversity or its speakers' authority has been strengthened. The proposed principle of mainstreaming without homogenization makes these concerns part of the evaluation of inclusion itself.

This principle does not treat all standardization as harmful. Shared conventions can support teaching, interoperability, and communication, and users may deliberately request standardized expression in a particular domain. The concern is the unexamined elimination of legitimate alternatives when a technological default is presented as neutral improvement. Evaluation should therefore distinguish transformations requested by users from transformations imposed by the system, and should examine whose preferences determine what counts as appropriate language.

Community authority is a corresponding governance condition. Cultural fidelity cannot be reduced to a model reproducing whatever a single evaluator identifies as authentic. Speakers may disagree over vocabulary, orthography, regional practices, or appropriate innovation. Grounding should preserve the possibility of such disagreement and provide mechanisms for contestation and correction. The framework thus treats internal diversity and decision rights as criteria for interpreting the quality of expansion, rather than assuming that higher usage automatically represents cultural benefit.

*P8. Among AI services producing comparable increases in access or use, those that preserve legitimate linguistic alternatives and community decision rights will exhibit stronger community-evaluated cultural fidelity and participatory control than those that impose unrequested standardization or external representational authority.*

Figure 1 presents the proposed relationships. Functional exclusion and homogenizing transformation are shown as interacting mechanisms rather than consecutive stages. Cultural grounding is positioned as an intervention directed at these mechanisms and at the conditions of meaningful use. The re-centering dimensions are parallel outcomes whose relationships require investigation. Table 1 distinguishes the explanatory process, the intervention, and the outcomes to be assessed.

Figure 1. Proposed mechanisms, grounding interventions, and cultural re-centering outcomes

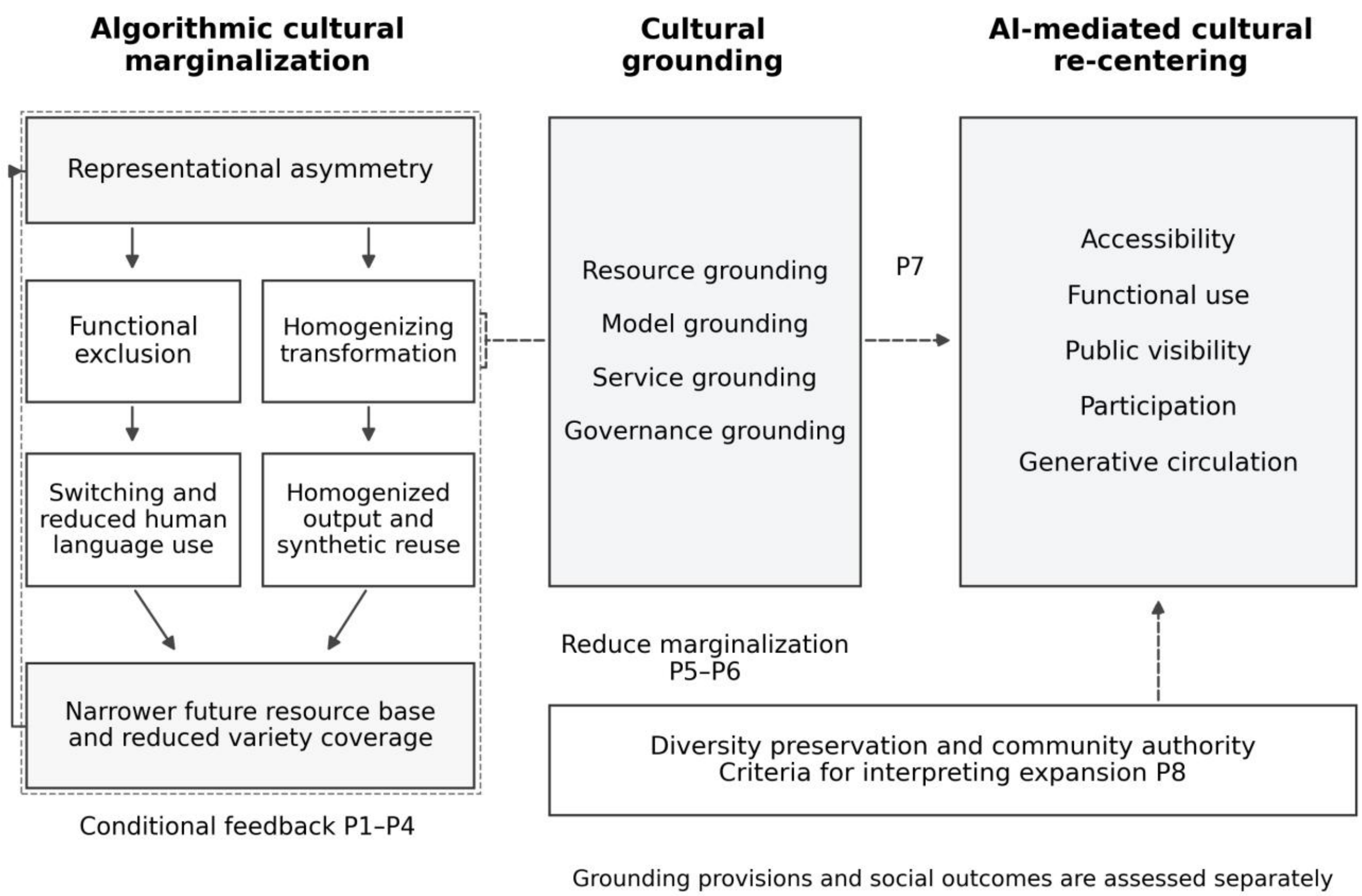

Note. Arrows indicate proposed relationships, not empirically established causal effects. Re-centering dimensions are parallel assessment domains rather than sequential stages.

Table 1. Analytical distinctions among the three constructs

| Construct | Definition | Analytical role | Evidence to assess |
|---|---|---|---|
| **Algorithmic** | Processes through which unequal | Explanatory | Performance gaps; |
| **cultural marginalization** | representation and AI-mediated practices weaken linguistic usability, distinctiveness, or knowledge circulation. | mechanisms | unrequested normalization; switching attributable to limitations; changes in resource coverage. |
| **Cultural grounding** | Design and governance provisions embedding linguistic variation, situated knowledge, provenance, task support, and community authority. | Sociotechnical intervention | Resource composition; model conditioning; implemented capabilities; enforceable decision and correction rights. |
| **AI-mediated cultural re-centering** | Observed expansion of the positions through which a marginalized language is encountered, used, and reproduced in digital society. | Social and ecosystem outcomes | Access; successful task use; visibility; participation; circulation, interpreted alongside diversity and authority. |

# 4 Hakka as an Illustrative Case and an Empirical Research Agenda

## 4.1 Historical language shift and contemporary AI environments

Hakka in Taiwan illustrates why the consequences of AI should be examined within a broader ecology of language use. Jan et al. (2016) identify associations among social context, parental exogamy, and Hakka language retention. Wang and Peng (2018), using national surveys conducted in 2004, 2011, and 2016, identify generational disruptions associated with television, migration from Hakka-concentrated settlements, and parental intermarriage. These findings locate language loss within transformations of media exposure, local linguistic concentration, and family arrangements.

AI may introduce another change in the environments through which speakers obtain knowledge and communicate. If comparable tasks consistently require less effort or yield more reliable outcomes in Mandarin Chinese or English than in Hakka, users may move toward those languages without rejecting Hakka identity or cultural value. This is a proposed mechanism of convenience-driven displacement, not an observed finding established by the present paper. Its investigation requires attention to language proficiency, existing multilingual practices, institutional expectations, and the kinds of tasks for which AI is used.

Hakka also demonstrates the importance of evaluating variation within a language label. Improvements in aggregate Hakka performance may conceal substantial differences among varieties, genres, and local contexts. A system can appear competent in familiar standardized tasks while failing to retain local vocabulary or culturally situated expression. The relevant questions consequently include whether requested varieties are maintained, whether alternative forms are accepted, and whether sources are interpreted in relation to their locality and historical context. The present framework does not rank Hakka varieties or claim that existing systems have produced measurable homogenization; those questions require appropriately stratified evidence.

The case is therefore analytically useful without serving as a demonstration of all proposed relationships. Historical research supports the importance of structural changes in language use. The effects of contemporary AI remain to be examined through designs that connect model behavior with actual user practices and resource circulation. Maintaining this distinction prevents a plausible analogy from being presented as an established causal account.

### 4.2 Evaluating linguistic transformation and cultural grounding

A first research program should examine how AI changes minority-language expression. Paired comparisons can use human-authored materials from multiple varieties, genres, and contexts, followed by rewriting or translation with several models. Prompts should specify the task and whether standardization is requested, so that appropriate adaptation can be distinguished from unrequested normalization. Original and transformed texts should be evaluated separately for semantic accuracy, linguistic acceptability, cultural-marker retention, and dialect fidelity. High semantic similarity should not be used as a substitute for the other dimensions.

Measures of lexical, syntactic, and semantic variation can provide useful baselines, as illustrated by Guo et al. (2024) and Sourati et al. (2026). Their use in Hakka would require language-appropriate validation, particularly for tokenization, orthographic variation, and the coverage of syntactic tools. Cultural and dialect markers should be identified with speakers familiar with the relevant varieties and contexts. Retention can be assessed as the proportion of applicable features preserved correctly, accompanied by qualitative analysis of substitutions, omissions, and changes in social meaning. Inputs without applicable markers should not be interpreted as perfect retention cases.

Cross-variety convergence provides another relevant outcome. Researchers can examine whether originally distinct varieties become more similar after transformation, while allowing for acceptable alternatives and task-specific conventions. Comparable ratings should use multiple evaluators and document agreement, disagreements, and the reasons for judgments. Changes in length, topic, genre, and requested audience should also be considered, because an apparent decline in variation may reflect legitimate changes in the task rather than homogenization alone.

A second research program should compare grounding interventions under controlled conditions. Useful contrasts include a general-purpose model, the same model with explicit linguistic instructions, generic retrieval, culturally curated retrieval, and retrieval designed to preserve relevant varieties and provenance. Keeping the base model, task set, and generation settings comparable would help isolate the contribution of particular interventions. Resource quality, retriever performance, and model instructions should be documented rather than combined into an unexplained label of cultural specialization.

Evaluation should include factual accuracy, source support, dialect fidelity, culturally consequential errors, stereotypical framing, and community judgments of situated appropriateness. Lertvittayakumjorn et al. (2025) provide a reason to evaluate factual knowledge and open-ended cultural familiarity separately. A useful intervention may improve one while leaving the other unchanged. Governance interventions also require independent observation:

decision rights, correction procedures, and whether accepted corrections change subsequent system behavior are more informative than the mere number of consultation meetings.

### 4.3 Connecting system performance with use and resource circulation

A third research program should examine whether improved support changes behavior. User studies should distinguish proficient speakers, heritage learners, and users with limited prior competence. Task-based observations can assess successful completion, time and effort, language switching attributable to system limitations, and subsequent willingness to continue using the language. Longitudinal follow-up is necessary to distinguish sustained changes from novelty effects. Expressed enthusiasm or intentions to learn can be informative, but they do not establish additional domains of actual use.

At the service and ecosystem levels, relevant evidence includes the range of tasks completed in the minority language, the distribution of users across varieties, the circulation of contributed materials, and the incorporation of minority-language sources into public-facing responses. Researchers should distinguish human-authored, collaboratively produced, and predominantly synthetic content, while acknowledging that these categories may be difficult to determine reliably. Content volume should be interpreted alongside diversity, provenance, and evidence of subsequent use.

Testing recursive relationships requires observations over time. New resource collections should be examined for changes in variety coverage, contributors, genres, and the influence of earlier AI outputs. Where consent and governance arrangements permit, evidence from interaction practices can be connected to subsequent contributions without assuming that all usage data should become training data. Increased interaction is not equivalent to authorization for collection or reuse. The proposed research agenda therefore links performance, behavior, and resource formation while treating community decision rights as part of the research design.

## 5 Discussion and Implications

### 5.1 Theoretical contribution and conceptual boundaries

The framework extends a domain-based account of language vitality to environments in which technological systems actively transform expression. Fishman (1991) provides an account of transmission and domains of use; Eisenlohr (2004) explains how mediation reconfigures communities; and Kornai (2013) distinguishes digital functionality from the survival of archived material. Algorithmic cultural marginalization brings these concerns into relation with model-driven reformulation, user adaptation, and the formation of future linguistic resources. Its contribution is this linkage across levels of analysis, rather than a claim that unequal language support or cultural normalization is newly discovered.

The separation of the three constructs is essential. Algorithmic cultural marginalization describes mechanisms through which disadvantage may be reproduced. Cultural grounding describes intentional design and governance provisions. Re-centering describes observed changes in a language's functional position. Resource coverage and implemented capabilities should consequently be measured independently of use and participation outcomes. Otherwise, an AI system would qualify as successful merely because the same features were counted both as its intervention and as its effects.

The framework also separates factual accuracy, linguistic accuracy, and cultural fidelity. This distinction connects research on unequal support with findings on homogenization and cultural representation (Agarwal et al., 2025; Ryan et al., 2024; Sourati et al., 2026; Wang et al., 2025). Its relevance is particularly clear where a fluent output preserves propositional content while erasing the distinctions through which speakers locate themselves socially and culturally. At the same time, fidelity must accommodate innovation and disagreement. A living language is not a fixed inventory of markers, and the evaluation of cultural change requires attention to speakers' intentions and the contexts in which new forms emerge.

## 5.2 Implications for development and language policy

The framework suggests evaluating minority-language AI as an interconnected infrastructure rather than a collection of isolated demonstrations. A speech recognizer, translation engine, chatbot, or archive may each provide value, but their combined contribution depends on whether users can move between tasks without repeated loss of linguistic control or source context. Development priorities should be informed by actual task requirements and by the varieties and user groups for which support remains weak.

Evaluation and procurement can incorporate these requirements without replacing conventional performance measures. Accuracy and usability remain necessary; they should be accompanied by stratified assessment across relevant varieties, checks on unrequested standardization, source-sensitive cultural evaluation, and functioning correction processes. Benchmark design should make visible whose performance improves and whose forms are treated as errors. It should also recognize that acceptable outputs may differ across tasks and communities, rather than imposing a single answer wherever alternatives are legitimate.

Resource policy should support human contributions and their contextual integrity. Diverse recordings, everyday genres, local knowledge, and contemporary creative work may be important precisely because they are poorly represented in readily available digital collections. Expanding access should be accompanied by agreed uses and mechanisms for correcting or withdrawing material where appropriate. Bird's (2020) critique of extractive language technology remains relevant when data collection is described as preservation but decision-making power remains external to the community.

For Hakka, the practical implication is to connect AI functionality with education, cultural production, and everyday interaction while assessing differential effects across varieties. This does not mean that every tool must address every domain. It means that individual projects should make their scope explicit and provide evidence of the linguistic practices they support. The strongest justification for investment is a demonstrated expansion of meaningful use under accountable governance, rather than the volume of generated content or the appearance of a language in a model's support list.

### 5.3 Limitations and conditions for further investigation

This paper develops a conceptual synthesis and does not provide empirical estimates of the proposed relationships. The cited studies address different languages, populations, and tasks; their findings cannot establish the effects of AI on Hakka without additional evidence. The propositions are therefore a research agenda rather than a validated explanatory model. Future studies may find that some mechanisms are weak, that interventions interact differently than expected, or that increased technological capability does not translate into sustained language use.

The framework's applicability also depends on sociolinguistic conditions. Languages differ in intergenerational transmission, institutional recognition, literacy practices, standardization, and the availability of digital resources. A system that expands access in one setting may have little effect in another because the binding constraint lies outside technology. Economic access, institutional language requirements, demographic changes, and existing digital platforms can influence outcomes independently of model quality. Comparative research should examine these conditions rather than assuming a uniform minority-language experience.

Community authority presents a further challenge because communities are not internally homogeneous. Evaluation arrangements may privilege formally recognized experts, dominant varieties, or participants with greater access to technology. Governance grounding should therefore be assessed for the distribution of decision rights and the opportunities available to less represented speakers. Disagreement over authenticity or standardization should be documented as part of the sociolinguistic setting, rather than treated automatically as an annotation problem to be eliminated.

Finally, AI functionality cannot substitute for interpersonal relationships or intergenerational transmission. It may expand opportunities for language use, but whether those opportunities become durable practices depends on families, communities, educators, institutions, and speakers themselves. The empirical task is to identify when technological support strengthens these relationships, when it merely adds another channel for dominant-language dependence, and how communities can retain authority over the linguistic futures that AI helps produce.